\documentclass[12pt,a4paper]{article}

\usepackage[truedimen,margin=30mm]{geometry}

\usepackage{mathrsfs}
\usepackage{amssymb}
\usepackage{amsmath}
\usepackage{ascmac}
\usepackage{amsthm}
\usepackage[dvipdfmx]{graphicx}
\usepackage{natbib}
\usepackage{setspace}
\usepackage{booktabs}

\usepackage{algorithm}
\usepackage{algpseudocode}

\usepackage{color}
\usepackage[unicode,colorlinks=true,allcolors=blue]{hyperref}

\usepackage{times}
\usepackage{comment}

\usepackage{titlesec}
\titleformat*{\section}{\large\bfseries}
\titleformat*{\subsection}{\it}

\newtheorem{thm}{Theorem}

\def\ep{{\varepsilon}}

\def\htau{{\widehat{\tau}}}
\def\hmu{{\widehat{\mu}}}
\def\hpi{{\widehat{\pi}}}

\def\bw{{\text{\boldmath $w$}}}

\title{{\bf Ensemble Doubly Robust Bayesian Inference via Regression Synthesis }}

\date{}

\begin{document}

\maketitle
\doublespacing

\vspace{-1.5cm}
\begin{center}
Kaoru Babasaki$^1$, Shonosuke Sugasawa$^2$, K\={o}saku Takanashi$^3$ and Kenichiro McAlinn$^4$

\medskip

\medskip
\noindent
$^1$Graduate School of Economics, Keio University\\
$^2$Faculty of Economics, Keio University\\
$^3$RIKEN Center for Advanced Intelligence Project\\
$^4$Department of Statistics, Operations, and Data Science, Fox School of Business, Temple University\\
\end{center}

\vspace{0.2cm}
\begin{center}
{\bf \large Abstract}
\end{center}

\vspace{-0cm}
The doubly robust estimator, which models both the propensity score and outcomes, is a popular approach to estimate the average treatment effect in the potential outcome setting.
The primary appeal of this estimator is its theoretical property, wherein the estimator achieves consistency as long as either the propensity score or outcomes is correctly specified.
In most applications, however, both are misspecified, leading to considerable bias that cannot be checked.
In this paper, we propose a Bayesian ensemble approach that synthesizes multiple models for both the propensity score and outcomes, which we call doubly robust Bayesian regression synthesis.
Our approach applies Bayesian updating to the ensemble model weights that adapt at the unit level, incorporating data heterogeneity, to significantly mitigate misspecification bias.
Theoretically, we show that our proposed approach is consistent regarding the estimation of both the propensity score and outcomes, ensuring that the doubly robust estimator is consistent, even if no single model is correctly specified.
An efficient algorithm for posterior computation facilitates the characterization of uncertainty regarding the treatment effect.
Our proposed approach is compared against standard and state-of-the-art methods through two comprehensive simulation studies, where we find that our approach is superior in all cases.
An empirical study on the impact of maternal smoking on birth weight highlights the practical applicability of our proposed method.


\section{Introduction}
\label{sec:org3a624a9}
Estimating the average treatment effect (ATE) of a binary treatment is one of the primary concerns in scientific research.
While randomized control trials (RCTs) are the gold standard, it is not always feasible or available, and considerable interest has been in developing methods to estimate the ATE from non-randomized trials or observational data.
In the potential outcome setting \citep{Rubin1974CausalEffects_JEducPsychol}, there are two ways to achieve RCT-like situations using non-RCT data; either by controlling for the propensity score \citep{Rosenbaum_Rubin1983PropensityScore_Biometrika} to emulate randomization, or by correctly estimating the (potential) outcomes.
Building on this idea, \citet{Robins1994DoublyRobust_JASA} proposed the augmented inverse probability weighted (AIPW) estimator, which simultaneously models both.
This method is also called the doubly robust (DR) estimator, since the estimate is consistent if either the propensity score model or outcome model is correctly specified \citep{Scharfstein_etal1999AdjustingForNonignorableDrop-OutRejoinder_JASA}.
The estimate is ``robust" in the sense that the researcher has two opportunities for valid inference.
This theoretical guarantee has made the DR estimator a popular approach for non-RCT data.

In practice, however, even one of these dual guarantees can be untenable.
More specifically, despite having two opportunities to correctly specify the models, complicated real-world data often leaves both misspecified.
Further complicating the issue is that there is no way to confirm or test whether either model is correctly specified.
Yet, when both are misspecified, it is known that the DR estimator has considerable bias \citep{Bang_Robins2005DoublyRobust_Biometrics}.
One prevalent approach to address this issue in current research is to select models based on theoretical considerations or some evaluation criteria, such as AIC/BIC or cross-validation.
However, both approaches impose strong assumptions on the data generating process, making the selection process itself, not robust.
As this instability of model selection directly affects the instability of the DR estimator, these strategies pose a significant practical challenge \citep{Robins2007DR_Performance_StatistSci}.

In response to this problem, we propose and develop a novel ensemble methodology for doubly robust estimation, which we call doubly robust Bayesian regression synthesis (BRS-DR).
Our proposed method is based on the Bayesian predictive synthesis framework \citep[BPS:][]{Mcalinn2019BPS_JoE}, from which we develop a method to synthesize different models for the propensity score and outcomes.
The synthesis functions we develop allow for adaptive weights, which are functions of the individual-level covariates, accounting for data heterogeneity via Bayesian updating.
Our development is guided by theoretical considerations, where we show that our synthesized estimates are consistent for both the propensity score and outcomes.
By robustifying each estimate through Bayesian synthesis, our doubly robust estimator is made further robust,
achieving consistency even when no single model is correctly specified.
Through extensive simulations, we confirm our theoretical justification, and show that BRS-DR outperforms existing methodologies.

This paper is organized as follows.
\hyperref[sec:orgf3a48cb]{Section 2} establishes the foundation, outlining the problem setting, basic concepts of DR estimation, and the application of the BRS framework within DR estimation.
\hyperref[sec:org5eb8d3d]{Section 3} develops the proposed method of BRS and theoretical support, along with a detailed discussion of its computational algorithm.
A two-part simulation study is presented in \hyperref[sec:org2c9d742]{Section 4}, where we test and validate our proposed methodology.
\hyperref[sec:orgeec5b04]{Section 5}
presents an empirical study investigating the impact of maternal smoking on birth weight, demonstrating the real-world applicability of our approach.
\hyperref[sec:org47e56a4]{Section 6} concludes the paper with a discussion of the essential findings and implications of our work.

\section{Ensemble Doubly Robust Estimation}
\label{sec:orgf3a48cb}

\subsection{Settings and assumptions}
\label{sec:org9dde731}

Consider a binary treatment \(Z \in \{0, 1\}\), with \(Z = 1\) indicating being treated and \(Z = 0\) indicating being untreated.
Let \(Y\) be the outcome of interest and \(X = (X_1, \dots, X_q)^{\top}\) be a set of \(q\)-dimensional covariates.
For \(n\) units, we denote the observed data by \(D = \{ (Y_i, X_i, Z_i) \}_{i=1}^n\).
Throughout this paper, we assume the stable unit treatment value assumption \citep[SUTVA,][]{Imbens_Rubin2015CausalInference}, which states that there is (i) no interference between units and (ii) no hidden variation of treatment.
We also assume the potential outcome framework \citep{Rubin2005CausalInferencePotentialOutcomes_JASA}, which states that we would observe \(Y_i(1)\) if unit \(i\) were treated and \(Y_i(0)\) if unit \(i\) were not treated.
Under these conditions, the observed outcome \(Y_i\) for unit \(i\) is defined as
\begin{equation}
\label{eq:observed_outcome}
Y_i = Y_i(1) Z_i + Y_i(0) (1 - Z_i).
\end{equation}

Our interest is in estimating the average treatment effect (ATE), given the treatment $T$, defined as
\begin{equation}
\label{eq:ate}
\tau = E[Y_i(1) - Y_i(0)].
\end{equation}
To identify the ATE, we assume the \emph{strong ignorability}
condition \citep{Rosenbaum_Rubin1983PropensityScore_Biometrika}, which implies
\emph{unconfoundedness}, $(Y_i(1), Y_i(0)) \perp Z_i | X_i$,  and \emph{positivity}, $0 < P(Z_i = 1 | X_i) < 1$.
Under these assumptions, there are two channels to estimate the ATE; through modeling the propensity score or the outcomes.
While a plethora of methods have been developed for both channels, as model-based inferences, they are susceptible to bias arising from model misspecification.

\subsection{Doubly robust estimator}

The double robust (DR) estimator, also known as the augmented inverse probability weighted estimator \citep{Robins1994DoublyRobust_JASA,Scharfstein_etal1999AdjustingForNonignorableDrop-OutRejoinder_JASA,Robins2007DR_Performance_StatistSci}, is a method to address the model misspecification issue by simultaneously modeling both the propensity score and outcomes.
The DR estimator is defined as
\begin{equation}
\label{eq:dr-estimator}
\htau_{\rm DR}
= \frac{1}{n} \sum_{i=1}^n
\left\{ \left[ \frac{Z_i Y_i}{\hpi(X_i)}
-\frac{ Z_i - \hpi(X_i)}{\hpi(X_i)}
\hmu_1(X_i) \right]
-\left[ \frac{(1 - Z_i) Y_i}{1 - \hpi(X_i)}
+\frac{ Z_i - \hpi(X_i)}{1 - \hpi(X_i)}
\hmu_0(X_i) \right] \right\},
\end{equation}
where $\hmu_k(X_i)$ is the estimator of $E[Y_i|X_i,Z_i=k]$ for $k=0,1$.
The reason why the DR estimator addresses the model misspecification issue is that it theoretically guarantees consistent estimates as long as either model for the propensity score or outcomes are correctly specified \citep{Scharfstein_etal1999AdjustingForNonignorableDrop-OutRejoinder_JASA}.
In this sense, it is ``doubly robust," as the estimator can fail to correctly specify one or the other and still be consistent.
Additionally, it reaches the semiparametric efficiency bound when the propensity score and the outcomes are both correctly specified.

While the DR estimator is robust if at least one is correctly specified, it is well-recognized that its performance breaks down considerably when both models are misspecified \citep{Bang_Robins2005DoublyRobust_Biometrics}; a more-than-common occurrence in real-world data.
Since the true propensity score or potential outcome is never observed in observational studies, this model misspecification issue is problematic for DR estimation, and there is no consensus on what is the best way-- either model selection or combination-- to mitigate this issue.

\subsection{Bayesian regression synthesis for outcomes and propensity scores}
\label{sec:org5eb8d3d}

To mitigate the model misspecification bias in DR estimation, we propose synthesizing multiple candidate models through covariate-dependent weights.
By synthesizing several estimates-- for both the propensity score and outcomes-- the aim is to leverage available information to build estimates that perform better than any one model (that is likely misspecified).
We also recognize that the misspecification of any model will not be homogeneous.
For example, a propensity score model might be correctly specified for some subgroups of the data, but not in others.
Our proposed method takes into account this heterogeneity by allowing weights to be adaptive to the covariates.
In this section, we will describe the Bayesian predictive synthesis framework through the development of the synthesis function for the outcomes, then develop the synthesis function for the propensity score as an extension.
Combining the two synthesis functions yields our proposed DR estimator (Section~\ref{sec:dr_brs}).
The proposed synthesis functions are developed based on their theoretical properties, described later in Section~\ref{sec:theory}, to not disturb the flow of the paper.

Suppose that we have $J$ models for either the propensity score or outcomes, which can include, but are not limited to, simple linear regressions, additive models, and nonparametric methods.
The only requirement for the $J$ models is that they provide some uncertainty quantification of its estimate, such as the posterior,  the bootstrap distribution, or some approximation.
For $j=1,\ldots,J$, denote $h_j(f_j(X)))$ as the estimate distribution from the $j$th model, and $f_j(X)$ be the random variable following the estimate distribution.
Regarding the outcome model, $h_j(f_j(X)))=h_j(\mu_j({X}))$, and for the propensity score model, $h_j(f_j(X)))=h_j(\pi_j({X}))$, thus $h_j(f_j(X)))$ is generalizing the two estimates of interest.

The general form of the Bayesian posterior distribution, given the $J$ estimates, can be expressed using the Bayesian predictive synthesis framework \citep{Mcalinn2019BPS_JoE}, as
\begin{equation}
\label{eq:bps}
p(Y | \mathcal{H}, X)
= \int \alpha(Y | f_1(X),\ldots,f_J(X))
\prod_{j=1:J} h_j(f_j(X)))df_j(X),
\end{equation}
where $\alpha(Y | f_1(X),\ldots,f_J(X))$ is the synthesis function that determines the form of posterior predictive, which may depend on unknown parameters.
This form of Bayesian updating (\ref{eq:bps}), originates in agent opinion analysis \citep{Genest&Schervish1985BayesianUpdating_AnnStat,West_Crosse1992AgentOpinion_JRSSB,West1992AgentForecast_JRSSB}, and is shown to be valid Bayesian posteriors.
Other than this paper, this form has been adopted for time series prediction \citep{McAlinn2020BPSMacroForecasting_JASA,kobayashi2023clustering} and spatial prediction \citep{Cabel2023BSPS}.
Recently, it has also been adopted for heterogeneous treatment effects \citep{sugasawa2023bayesianCausalSynthesis}, where several conditional average treatment effects are synthesized using a specific synthesis function.
BPS for regression contexts, either for the propensity score or outcomes, has not been done, let alone in the context of DR estimation.

Let us now consider extending the BPS framework for the regression of the outcomes.
Given $\mu_1(X),\ldots,\mu_J(X)$, we employ a synthesis model, $$\alpha(Y | \mu_1(X),\ldots,\mu_J(X))=\phi(Y;  \beta_0(X) + \sum_{j=1}^J \beta_j(X) \mu_j(X), \sigma^2),$$ where $\beta_0(X),\ldots,\beta_J(X)$ are unknown functions and $\sigma^2$ is a unknown variance parameter.
Under the specification, the posterior (\ref{eq:bps}) can be regarded as the posterior predictive distribution derived from the following hierarchical model:
\begin{equation*}
Y = \beta_0(X) + \sum_{j=1}^J \beta_j(X) \mu_j(X) + \ep,
\quad \ep \sim N(0, \sigma^2), \ \ \ \mu_j(X)\sim h_j.
\end{equation*}
To complete the specifics of the above model,
we assume that the unknown regression function \(\beta_j(X)\) follows an isotropic Gaussian process.
In particular, to assure computational stability under large sample sizes,
we employ a nearest-neighbor Gaussian process \citep{Datta2016HierachicalNNGP_JASA},
that is,
\(\beta_j(X) \sim \mathrm{NNGP}_m( \bar{\beta}_j(X), \tau_j^2 C(\cdot ; \psi_j))\),
where \(\tau_j^2\) is an unknown variance parameter,
\(\psi_j\) is an unknown range parameter,
\(C(\cdot; \psi)\) is a valid correlation function with spatial range parameter \(\psi\),
and \(m\) is the number of nearest neighbors.
For each \(j = 1, \dots, J\),
\(\bar{\beta}_j\) represents a global weight for the \(j\)-th method
and also serves as a prior mean for the heterogeneous weight \(\beta_j(X)\).
Although it is an option to assign an informative prior to \(\bar{\beta}_j\),
we opt for a default approach where \(\bar{\beta}_0\) is set to \(0\) and \(\bar{\beta}_j\) is set to \(1/J\) for \(j=1, \dots, J\).
This default choice results in the prior synthesis function representing an equal-weight average of the \(J\) estimators.

Given observed samples, $\{(Y_i, X_i)\}_{ i = 1, \dots, n }$, the model can be expressed as
\begin{equation}
\label{eq:brs_gauss}
\begin{split}
&Y_i = \beta_0(X_i) + \sum_{j=1}^J \beta_j(X_i) \mu_j(X_i) + \epsilon_i, \quad \epsilon_i \sim N(0, \sigma^2),\\
&(\beta_j(X_1), \dots, \beta_j(X_n)) \sim N( \bar{\beta}_j, \tau_j^2 H_m(\psi_j; \mathcal{X}) ), \ \ \ j=1,\ldots,J,
\end{split}
\end{equation}
where \(H_m(\psi_j; \mathcal{X})\) represents the \(n \times n\) covariance matrix of the joint distribution for \(n\) observations based on an \(m\)-nearest neighbor Gaussian process on the space \(\mathcal{X}\), representing the domain of \(X_i\).
For the unknown parameters, \(\theta = (\sigma^2, \tau_0^2, \dots, \tau_J^2, \psi_0, \dots, \psi_J)\), we implement priors, \(\sigma^2 \sim \mathrm{IG}(\delta_\sigma/2, \xi_\sigma/2)\),
\(\tau_j^2 \sim \mathrm{IG}(\delta_j/2, \xi_j/2)\), and \(\psi_j\sim U(\underline{c}_{\beta}, \bar{c}_{\beta})\).
Then, the posterior distribution of $\theta$ as well as the covariate-dependent weight  \(\beta_0(\cdot), \dots, \beta_J(\cdot)\),
can be approximated using a Markov chain Monte Carlo (MCMC) algorithm.
The detailed algorithm is given in the Supplementary Material.

To synthesize multiple models for propensity score, we consider BRS for binary response.
Let $Z_i$ be the binary response and, for $j=1,\ldots,J$, we have $h_j(\pi_j(X))$, the $J$ number of estimate distributions of the propensity score.
Our proposal is the following synthesis function:
\begin{equation}
\label{eq:brs_binary}
Z_i \sim \mathrm{Ber} \left( \frac{\exp(\eta_i)}{1+ \exp(\eta_i)} \right),
\quad \eta_i = \beta_0(X_i) + \sum_{j=1}^J \beta_j(X_i) \pi_j(X_i).
\end{equation}
Similar to (\ref{eq:brs_gauss}), we assign a nearest-neighbor Gaussian process on $\beta_j(X_i)$ and the same priors on the unknown parameters.
Using the Polya-gamma data augmentation, the posterior predictive distribution under the synthesis model (\ref{eq:brs_binary}) can be efficiently approximated by a Markov chain Monte Carlo algorithm, where the details are given in the Supplementary Material.

\subsection{Doubly robust posterior distribution of ATE\label{sec:dr_brs}}
Given the posterior distribution of the outcome regression, $\mu_1(X_i)$ and $\mu_0(X_i)$, and propensity score, $\pi(X_i)$, we obtain the posterior distribution of the ATE.
First, we define the conditional posterior of the ATE given $(\mu_1(X_i),\mu_0(X_i),\pi(X_i))$ through Bayesian bootstrap \citep{Rubin1981BayesianBootstrap_AnnStat}.
Let $\bw=(w_1,\ldots,w_n)$ be a vector of random weights generated from ${\rm Dir}(1,\ldots,1)$ and define
\begin{align*}
\tau_w(\mu_1,\mu_0,\pi)
 = \sum_{i=1}^n w_i\bigg\{ 
 &\left[ \frac{Z_i Y_i}{\pi(X_i)}-\frac{ Z_i - \pi(X_i)}{\pi(X_i)}\mu_1(X_i) \right]\\
& \ \ \ \ \ 
-\left[ \frac{(1 - Z_i) Y_i}{1 - \pi(X_i)}
-\frac{ Z_i - \pi(X_i)}{1 - \pi(X_i)}
\mu_0(X_i) \right] \bigg\},
\end{align*}
where $\mu_k=(\mu_k(X_1),\ldots,\mu_k(X_n))$ for $k=1,2$ and $\pi=(\pi(X_1),\ldots,\pi(X_n))$.
Note that $E[w_i]=1/n$ and the expectation of $\tau_w(\mu_1,\mu_0,\pi)$ with respect to $\bw$ is the standard doubly robust estimator obtained by replacing $w_i$ with $1/n$, which is denoted by  $\tau(\mu_1,\mu_0,\pi)$.
The general theory of the Bayesian bootstrap \citep{Rubin1981BayesianBootstrap_AnnStat,Lyddon_etal2019GeneralBayesianUpdatingLoss-LikelihoodBootstrap_Biometrika} suggests that the asymptotic distribution of $\sqrt{n}\{\tau_w(\mu_1,\mu_0,\pi)-\tau(\mu_1,\mu_0,\pi)\}/V(\mu_1,\mu_0,\pi)^{1/2}$ given the observed data $\mathcal{D}_n$ under $n\to\infty$ is the standard normal distribution, where
\begin{align*}
V(\mu_1,\mu_0,\pi)=\frac1n\sum_{i=1}^n \bigg\{ & \left[ \frac{Z_i Y_i}{\pi(X_i)}-\frac{ Z_i - \pi(X_i)}{\pi(X_i)}\mu_1(X_i) \right]\\
&-\left[ \frac{(1 - Z_i) Y_i}{1 - \pi(X_i)}
-\frac{ Z_i - \pi(X_i)}{1 - \pi(X_i)}
\mu_0(X_i) \right] \bigg\}^2.
\end{align*}
Hence, the distribution of $\tau_w(\mu_1,\mu_0,\pi)$ has frequentist justification under large samples, and can be regarded as an approximate posterior distribution of $\tau$ given $(\mu_1,\mu_0,\pi)$.
When either of $(\mu_1, \mu_0)$ or $\pi$ is correctly specified, the posterior (bootstrap) mean $\tau(\mu_1,\mu_0,\pi)$ converges to the true value of $\tau$ under $n\to\infty$.

Finally, the marginal posterior of $\tau$ can be obtained by using the posterior samples of $(\mu_1,\mu_0,\pi)$, which we call doubly robust posterior via Bayesian regression synthesis (BRS-DR).
The step-by-step procedures to generate random samples from BRS-DR are summarized in Algorithm~1.

\begin{algorithm}[H]
\label{algo:bb}
\caption{(BRS-DR) Doubly robust posterior with Bayesian regression synthesis}

\medskip
\begin{algorithmic}[1]
\State
Prepare prediction means and variances of $J$ models for outcome regression and propensity score.
\State
Run Markov chain Monte Carlo algorithm of Bayesian regression synthesis for both outcome regression and propensity scores to obtain $B$ posterior samples, $(\mu_1^{(b)}(X_i), \mu_0^{(b)}(X_i), \pi^{(b)}(X_i))$ for $b=1,\ldots,B$ and $i=1,\ldots,n$.

\For{\(b = 1, \dots, B\)}
    \State Draw random weights $(w_1^{(b)}, \dots, w_{n}^{(b)}) \sim \text{Dir}(1, \dots, 1)$.
    \State Compute the weighted average:
    \begin{align*}
    \tau^{(b)}
    = \sum_{i=1}^n w_{i}^{(b)}
    \bigg\{ &\left[ \frac{Z_i Y_i}{\pi^{(b)}(X_i)}
    -\frac{ Z_i - \pi^{(b)}(X_i)}{\pi^{(b)}(X_i)}
    \mu_{1}^{(b)}(X_i) \right]\\
    & \ \ \ \  \ 
    -\left[ \frac{(1 - Z_i) Y_i}{1 - \pi^{(b)}(X_i)}
    +\frac{ Z_i - \pi^{(b)}(X_i)}{1 - \pi^{(b)}(X_i)}
    \mu_{0}^{(b)}(X_i) \right] \bigg\}
    \end{align*}
\EndFor
\State Output the posterior samples $\{\tau^{(1)},\ldots,\tau^{(B)}\}$.
\end{algorithmic}
\end{algorithm}

Based on the posterior samples of $\tau$, we can compute a point estimate by the posterior mean and an interval estimate based on the $95\%$ credible interval.
Since the posterior of $(\mu_1,\mu_0,\pi)$ can successfully capture uncertainty in the estimation of outcome regression and propensity score, the final posterior of $\tau$ can also take into account the estimation and inference uncertainty of the ATE.

\subsection{Theoretical properties of BRS\label{sec:theory}}

In this section, we provide the theoretical
justification for our proposed BRS-DR. Specifically, we derive its asymptotic behavior
regarding its consistency in terms of estimating the propensity score and outcome.

We first consider consistency of propensity scores. 
Let $\pi_j(X)$ be the predictive distribution of the logit-transformed propensity score based on the $j$th model.
Given a vector of binary treatment, $Z_1,\ldots,Z_n$, BRS predicts the propensity score based on the model (\ref{eq:brs_binary}).
Here we assume that the true distribution, $p^{\ast}(Z|x)$, of the treatment given covariates $X$ is within the BRS predictive model.
In other words, there exists at least one,
$\delta_{0}^{*}(x)$ and $\delta_j^{*}(x)$, and the following holds:
\begin{alignat*}{1}
p^{*}(Z \mid x)
& =\frac{
\exp\Big\{Z\big(\delta_{0}^{*}(x)+\sum_{j=1}^{J}\delta_{j}^{*}(x)\pi_j(x)\big)\Big\}}{1+\exp\big(\delta_{0}^{*}(x)+\sum_{j=1}^{J}\delta_{j}^{*}(x)\pi_{j}(x)\big)}.
\end{alignat*}
Let $\pi_{\ast}(x)$ be the posterior predictive expectation of the above model, that is, $\pi_{\ast}(x)=P(Z=1 | x)$.
Our goal is to construct a consistent estimator of $\pi_{\ast}(x)$ for arbitrary $x$.
We define $\hat{\pi}\left(x\right)$ to be the posterior predictive mean based on the BRS model.
Then, we have the following property, where the roof is given in the Supplementary Material:

\begin{thm}\label{thm:consistency}
The synthesized propensity score estimate $\hat{\pi}(x)$ is consistent, namely, $\hat{\pi}(x)\overset{\textrm{p}}{\rightarrow}\pi_{\ast}\left(x\right)$.
\end{thm}

We next consider consistency of outcome models.
Let $\mu_j(X_i)$ be the predictive distribution of the outcome based on the $j$th model.
Given $\mu_1(X_i),\ldots,\mu_J(X_i)$, BRS for outcomes predicts the regression mean with the Gaussian model~(\ref{eq:brs_gauss}).
We assume that the true distribution, $p^{\ast}(Y|X)$, of the outcome $Y$ given covariates $X$ is within the BRS predictive model, that is, there exists at least one $\beta_{0}^{*},\beta_{1}^{*},\cdots,\beta_{J}^{*}$ such that the DGP can be expressed as,
\begin{alignat*}{1}
Y & =\beta_{0}^{*}\left(X\right)+\sum_{j=1}^{J}\beta_{j}^{*}(X)\mu_{j}(X)+\varepsilon,\  \ \ \varepsilon\sim N \left(0,\sigma^{2}\right).
\end{alignat*}
The estimand is the expectation of the above true distribution given $X$, denoted by $m_{\ast}(X)$.
Also, we define $\hat{m}(X)$ to be the posterior predictive expectation of the posterior predictive distribution based on the BRS model (\ref{eq:brs_gauss}).
Then, we have the following property:

\begin{thm}\label{thm:Consistency-Predic}
Assume that $\beta_{0}^{*},\beta_{1}^{*},\cdots,\beta_{J}^{*}$ are in $L^{2}\left(X\right)$, and the parameter space, $L^{2}\left(X\right)$, is complete and separable.
Then, we have $\hat{m}(X)\overset{\textrm{p}}{\rightarrow} m_{\ast}(X)$.
\end{thm}

The proof is identical to the proof of Theorem~\ref{thm:consistency}, and is thus omitted.

\section{Simulation Study}
\label{sec:org2c9d742}
We compare the performance of our proposed BRS-DR method with existing doubly robust estimators through simulation studies.
To emulate real-world conditions, both propensity and outcomes exhibit complex functional forms.
Under such situations, relatively simple statistical models are likely to be misspecified.
Throughout this section, let $X=(X_1,\ldots,X_q)$ be a set of covariates, $Z$ be a binary treatment indicator and $Y$ be an outcome.
We consider the following data generating process of $Y$ and $Z$:
\begin{equation}\label{DGP}
Y=\mu(X, Z) + \ep, \ \ \ \ \ep\sim N(0,\sigma^2), \ \ \ \ Z\sim {\rm Ber}\left(\frac{e^{\eta(X)}}{1+e^{\eta(X)}}\right),
\end{equation}
where $\mu(X, Z)$ and $\eta(X)$ are some functions of $X$ and $Z$.
In what follows, we adopt two scenarios for the true structures of $\mu(X, Z)$ and $\eta(X)$.

\subsection{Simulation 1: complex functional forms}
\label{sec:sim1}
Following \citet{Kang_Schafer2007DoublyRobust_StatistSci}, we consider the following structures:
\begin{align*}
\mu(X, Z) &=210 + 27.4 U_1(X) + 13.7 U_2(X) + 13.7 U_3(X) + 13.7 U_4(X), \\
\eta(X) &=- U_1(X) + 0.5 U_2(X) - 0.25 U_3(X) - 0.1 U_4(X)
\end{align*}
with $\sigma^2=1$ in (\ref{DGP}).
Here $U_1(X),\ldots,U_4(X)$ are independently generated from $N(0,1)$ and are implicit functions of $X_1,\ldots,X_4$ through $X_1= \exp( U_1/2)$, $X_2=U_2/(1 +e^{U_1})+ 10$, $X_3=(U_1 U_3/25 + 0.6)^3$ and $X_4=(U_2 + U_4 +20)^2$.
Hence, $U_1,\ldots,U_4$ are unobserved confounders and are not observed, and only $X_1,\ldots,X_4$ are observed.

Under this setting, the true treatment effects for all the individuals are $0$, so the true ATE is also $0$.
Based on the simulated dataset with three different sample sizes, \(n = 200, 1000, 2000\), we estimate the ATE using the following methods:
\begin{itemize}
\item[-]
{\bf Generalized linear model (GLM)}: DR estimator using generalized linear models (logistic regression for propensity score and linear regression for outcome) with covariates $X_1,\ldots,X_4$.

\item[-]
{\bf Generalized quadratic model (GQM)}: DR estimator using generalized linear models (logistic regression for propensity score and linear regression for outcome) with covariates $X_1,\ldots,X_4$ and $X_1^2,\ldots,X_4^2$.

\item[-]
{\bf Generalized additive model (GAM)}: DR estimator using generalized additive models \citep{Hastie_Tibshirani1986GAM_StatistSci} for propensity score (logistic additive regression) and outcome (linear additive regression), implemented with the R package ``gam'' with default settings of tuning parameters.
\end{itemize}

Additionally, we synthesize the results of the above three models through the following ensemble methods:
\begin{itemize}
\item[-] {\bf Simple averaging (SA)}: DR estimator with equally weighted averages of GLM, GQM, and GAM for both propensity score and outcomes.

\item[-]
{\bf Smoothed AIC averaging (SIC)}: Using the normalized Akaike weights \citep[e.g.][]{Burnham_Anderson2002ModelSelectionAndMultimodelInference} defined by $w_j \propto \exp( - \mathrm{AIC}_j /2)$ for $j=1,2,3$ for the three models, GLM, GQM and GAM, compute DR estimator with weighted averages for both propensity score and outcomes.

\item[-]
{\bf Bayesian model averaging (BMA)}: Under uniform prior on the model probability, apply Bayesian model averaging to combine the three models for both propensity score and outcomes, and compute the DR estimator.

\item[-]
{\bf Bayesian regression synthesis (BRS)}: Applies the proposed BRS to synthesize the three models, where the latent factors are approximated normal distribution with means and variance corresponding to the point estimates and standard errors of the regression term.
The number of nearest neighbors is set to \(m = 10\) and 1500 posterior samples after discarding 500 burn-in samples are used to approximate the DR posterior.
\end{itemize}

Based on $R=500$ Monte Carlo replications, we computed bias and mean squared errors (MSE), defined as ${\rm Bias}=R^{-1}\sum_{r=1}^R(\htau^{(r)}-\tau)$ and ${\rm MSE}=R^{-1}\sum_{r=1}^R(\htau^{(r)}-\tau)^2$, where $\htau^{(r)}$ is the estimate of ATE under $r$th replication and $\tau$ is the true ATE value ($\tau=0$ in this case).
Furthermore, to evaluate the performance in terms of uncertainty quantification, we obtained $95\%$ confidence/credible intervals of $\tau$ and computed the coverage probability (CP) and average length (AL).

The results are reported in Table~\ref{tab:sim1}.
The results consistently show that BRS has the smallest absolute bias and MSE across different sample sizes.
For instance, at \(n = 2000\), BRS exhibits a markedly lower bias (-0.915) and MSE (1.087) compared to other methods.
This trend is consistent across smaller sample sizes, indicating BRS's superior accuracy in point estimation of the ATE.
Notably, as the sample size increases, the performance of BRS improves in both bias and MSE, a trend not observed with other methods.
This result suggests that BRS efficiently utilizes larger sample sizes to enhance accuracy.

Turning to interval estimation, which is assessed by CP and AL of confidence/credible intervals, BRS stands out for its high CP, close to the target 95\%.
Notably, the AL of BRS is relatively shorter or comparable to other methods, especially at larger sample sizes, suggesting that BRS does not overestimate the degree of uncertainty but accurately captures it.
Overall, the data from both tables highlight the efficacy of BRS-DR in accurately estimating the ATE, especially in terms of precision (lower bias and MSE) and uncertainty (CP closer to 95\%) across different sample sizes.

\begin{table}
\caption{Bias and squared root of mean squared error (RMSE) of point estimates and coverage probability (CP) and average length (AL) of $95\%$ confidence/credible intervals of the ATE, based on 500 Monte Carlo replications, under simulation study 1.
}
\label{tab:sim1}
\centering
\medskip
\begin{tabular}{cccccccccccccccc}
\hline
 & $n$ &  & GLM & GQM & GAM & SA & SIC & BMA & BRS \\
\hline
 & 200 &  & -7.51 & -29.41 & -3.93 & -5.09 & -4.17 & -4.17 & -2.36 \\
Bias & 1000 &  & -8.56 & -62.57 & -4.00 & -6.36 & -4.25 & -4.25 & -1.24 \\
 & 2000 &  & -9.03 & -103.49 & -4.05 & -6.63 & -4.33 & -4.33 & -0.88 \\
 \hline
 & 200 &  & 11.21 & 72.60 & 5.28 & 6.41 & 5.54 & 5.54 & 3.68 \\
RMSE & 1000 &  & 9.39 & 111.02 & 4.23 & 6.60 & 4.49 & 4.49 & 1.48 \\
 & 2000 &  & 9.58 & 162.75 & 4.14 & 6.76 & 4.45 & 4.45 & 1.01 \\
 \hline
 & 200 &  & 56.7 & 46.7 & 56.9 & 56.5 & 56.5 & 56.3 & 100.0 \\
CP (\%) & 1000 &  & 4.3 & 1.4 & 13.9 & 4.8 & 14.4 & 14.4 & 96.1 \\
 & 2000 &  & 0.8 & 0.0 & 2.0 & 0.3 & 2.8 & 2.3 & 88.2 \\
 \hline
 & 200 &  & 22.07 & 102.23 & 9.76 & 12.92 & 10.36 & 10.35 & 26.32 \\
AL & 1000 &  & 13.73 & 197.88 & 4.84 & 8.19 & 5.42 & 5.41 & 5.49 \\
 & 2000 &  & 11.86 & 329.27 & 3.48 & 6.60 & 4.11 & 4.11 & 3.04 \\
\hline
\end{tabular}
\end{table}

\subsection{Simulation 2: Existence of unobserved confounders}
\label{sec:orge185bfb}
Next, we consider scenarios where the treatment effect varies according to covariates, and there exist omitted confounding factors.
We consider the following data generating process inspired by
\citet{Yu_etal2023SemiparaBayesDoublyRobust_JStatPlanInf}:
\begin{align*}
\mu(Z, X) &= Z + 2 Z X_1 + X_1 + X_2 + X_3 + X_4 + 0.25X_1^2 + 0.75 X_2 X_4 + 0.75 X_3 X_4 \\
\eta(X)&=0.3 X_1 + 0.9 X_2 - 1.25 X_3 + 1.5 X_4,
\end{align*}
with $\sigma^2=1$ in (\ref{DGP}), where $X_1, X_2 \sim N(1,1)$, $X_3, X_4 \sim N(-1,1)$ and $X_5, X_6, \dots, X_q \sim N(0,1)$.
Note that the HTE given covariates is calculated as \(\mu(1,X) - \mu(0,X) = 1 + 2 X_1\), with the true ATE deduced to be \(\mathbb{E}[1 + 2 X_1] = 3\).
Our analysis first observes the case with full covariate observance, then proceeds to a scenario where \(X_3\) is omitted, illustrating the effects of omitted variable bias.
We examine six distinct combinations of sample sizes \(n = 200, 1000, 2000\) and the number of covariates $q=4$ and $100$.
500 Monte Carlo replications are generated and analyzed for the six scenarios as in the previous simulations.
The methods chosen for the ATE estimation align with previous simulations and encompass the GLM, GQM, GAM, and ensemble methods, such as SA, SIC, BMA, and BRS, as considered in the previous section.
Given the high-dimensional nature of the data when \(q=100\), variable selection is needed.
The Lasso technique is applied via the ``glmnet'' package in R for GLM and GQM.
For GAM, the ``SFGAM'' function within the ``sparseGAM'' package is employed.
The regularization parameter \(\lambda\) is determined by cross-validation techniques, ``cv.glmnet'' for Lasso and ``cv.SFGAM'' for GAM, to ensure the most stable and feasible estimation under high-dimensional settings.

\begin{figure}[htbp]
\includegraphics[width=0.95\textwidth]{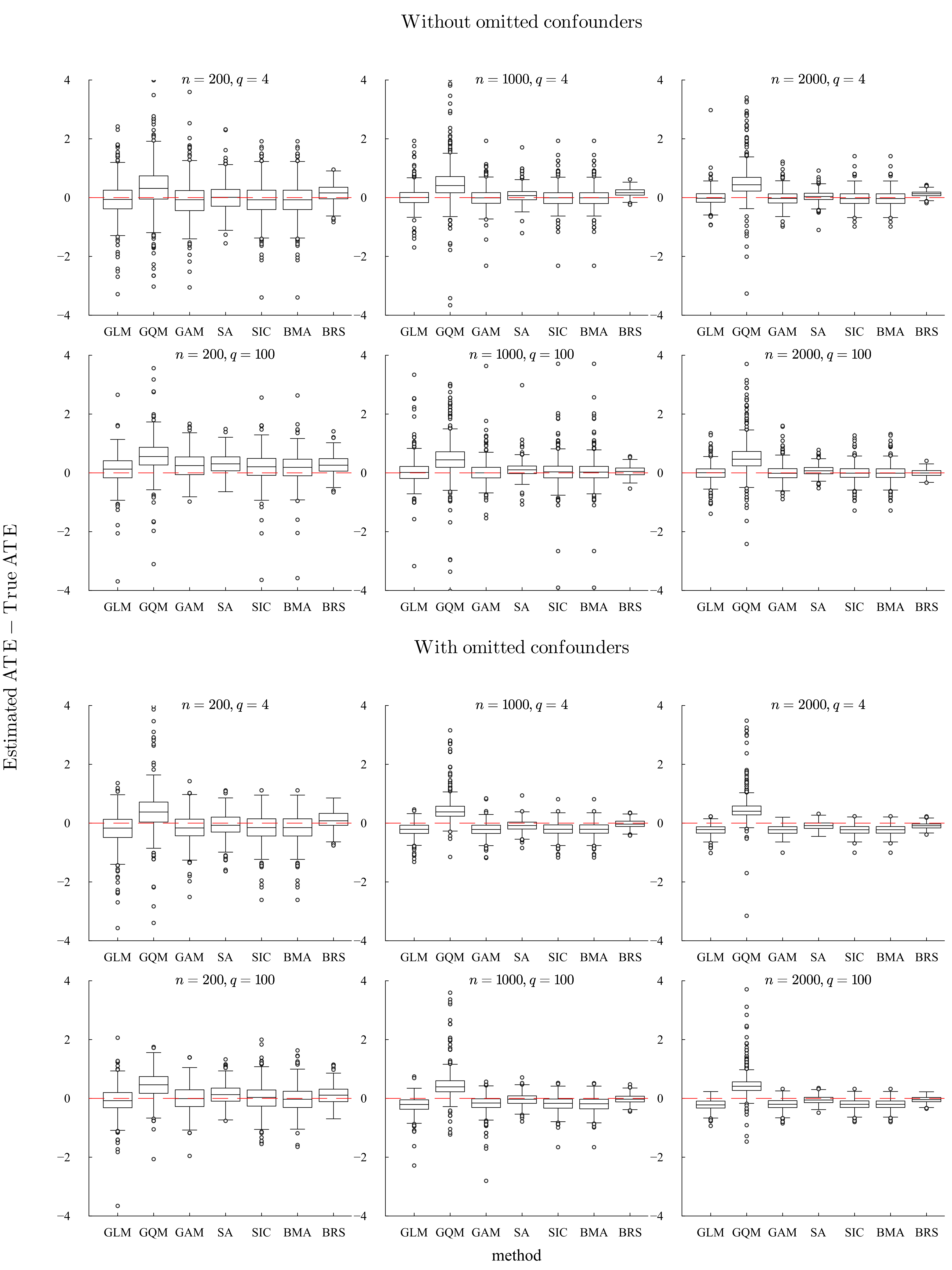}
\caption{\label{fig:sim2:errorbox}
Errors of point estimates of the ATE for 500 replications (with and without omitted confounders) under Simulation 2}
\end{figure}

The boxplots of the errors of point estimates of the ATE for 500 replications are shown in Figure \ref{fig:sim2:errorbox}.
This figure confirms the superiority of BRS across various sample sizes (denoted by \(n\)) and the number of covariates (denoted by \(q\)).
When \(X_3\) is observed and \(q=4\) (first row of Figure \ref{fig:sim2:errorbox}), the estimates of BRS are slightly positively biased compared to other methods, but the variance is much smaller.
BRS consistently displays reduced bias and variance across all scenarios, with the smallest MSE in all cases.
Especially when \(X_3\) is omitted, the performance of BRS seems to be the only method that accurately captures the ATE with slight bias and variance.
Notably, Figure \ref{fig:sim2:errorbox} also suggests that BRS yields fewer outliers, indicating its stability and robustness.
These results suggest that BRS-DR is a robust and accurate method for estimating the ATE, even in scenarios with heterogeneous treatment effects and omitted confounders.

\section{Empirical Study: The Effect of Maternal Smoking on Birth Weight}
\label{sec:orgeec5b04}
To empirically evaluate our proposed method, we apply it to real-world data from a widely used dataset given in \citet{Cattaneo2010SemiparametricEstOfMultiValuedTreatment_JoE}.
The dataset is a subset of the birthweight dataset originally used in \citet{Almond_etal2005CostsOfLowBirthWeight_QJE}, which concluded that maternal smoking during pregnancy has a considerable negative effect on birth weight.
It provides 4,642 detailed records on parents and birth information for singletons born in Pennsylvania.
Following previous studies, we focus on estimating the ATE of maternal smoking on birth weight by using different doubly robust estimators, including our proposed method.
The dataset includes 10 covariates, mother's age, mother's education level, father's education level, number of prenatal care visits, months since last birth, previous births where a newborn died, mother's marital status ('1' for married and '0' for unmarried), mother's alcohol consumption ('1' for consumption and '0' for none), mother's race ('1' for white and '0' for non-white) and first baby ('1' for yes and '0' for no), where further details are given in \citep{Cattaneo2010SemiparametricEstOfMultiValuedTreatment_JoE}.
The data characteristics show that the groups differ significantly across various covariates, which highlights potential challenges for estimating the causal effect of maternal smoking on birth weight.
On average, newborns in the smoking group had a
lower birth weight (3137.66 grams) compared to the non-smoking group (3412.91 grams), suggesting a negative impact of maternal smoking as in \citet{Almond_etal2005CostsOfLowBirthWeight_QJE}.
However, the smoking group had a significantly lower average age, education level, and number of prenatal care visits, and a higher average number of months since the last birth, number of previous births where the newborn died, and alcohol consumption.
In addition, the smoking group had a significantly higher proportion of unmarried mothers, non-white mothers, and mothers with their first baby.
These imbalances highlight the potential confounding effects of the covariates on the potential outcomes, which emphasizes the importance of using a statistical method that can adjust for the covariates.

We estimate the ATE of maternal smoking on birth weight by using seven doubly robust estimators, as considered in Section~\ref{sec:sim1}.
Table \(\ref{tab:BwResults}\) shows the estimates of the ATE of maternal smoking on birth weight and the corresponding standard errors,
95\% confidence/credible intervals and lengths for the seven DR estimators.

\begin{table}
\centering
\caption{Estimates of the ATE of Maternal Smoking on Birth Weight}
\begin{tabular}{lcccc}
\hline
\multicolumn{1}{c}{Method} & \multicolumn{1}{c}{Est} & \multicolumn{1}{c}{SE} & \multicolumn{1}{c}{95\% CI} & \multicolumn{1}{c}{CI Length}  \\ \hline
GLM & -231.13 & 24.80 & [-279.85, -182.02] & 97.83 \\
GQM & -230.26 & 34.93 & [-304.73, -166.92] & 137.81 \\
GAM & -233.52 & 33.88 & [-303.54, -170.98] & 132.56 \\
SA & -229.90 & 28.24 & [-287.06, -176.53] & 110.53 \\
SIC & -233.81 & 34.14 & [-305.77, -171.42] & 134.35 \\
BMA & -233.47 & 34.24 & [-302.66, -168.88] & 133.77 \\
BRS & -219.02 & 19.74 & [-256.05, -180.82] & 75.23 \\ \hline
\end{tabular}
\label{tab:BwResults}
\end{table}

The results in Table \(\ref{tab:BwResults}\) reveal several noteworthy characteristics of the estimated ATE of maternal smoking on birth weight.
All methods consistently indicate a negative effect of maternal smoking, with point estimates ranging from -219.02 to -233.81 grams.
BRS is particularly notable for its largest estimate of -219.02 grams, which significantly differs from the naive estimate of -275.3 grams (derived from simple group averages).
This observation hints at the potential strength of BRS in mitigating confounding bias.
BRS also distinguishes itself by demonstrating the smallest standard error (19.74) and the narrowest 95\% confidence interval (75.23). At the same time, other ensemble methods (SA, SIC, and BMA) are just averages of LM, QM, and GAM.
This result implies that BRS evaluates uncertainty through its probabilistic model instead of simply averaging uncertainties.
This methodological distinction subtly underscores the robustness of the BRS method, particularly in its precise quantification of uncertainty, which is a critical aspect of statistical analysis.

\section{Discussion}
\label{sec:org47e56a4}
This paper focused on creating a new doubly robust (DR) estimator for the average treatment effect (ATE) by utilizing the Bayesian Predictive Synthesis (BPS) framework in the context of propensity score and outcome modelings.
The study also aimed to demonstrate the efficacy of the proposed method through a series of simulations and explore its potential to overcome the limitations of existing DR estimators.
Previous research on ensemble approaches to the DR estimation of the ATE is limited, and our study is the first to apply the BPS framework in this context.
The findings suggest that the BPS framework could offer a novel solution to the limitations faced by current DR estimators.

Our proposed method, Bayesian Regression Synthesis for Doubly Robust Estimation (BRS-DR), demonstrated superior performance in estimating the ATE across various simulation settings, particularly in scenarios where other DR estimators fail to perform well due to issues like missing variables, high sparsity, and complexity of the data structure.
Additionally, BRS-DR showed high accuracy in assessing the uncertainty of the estimated ATE, a crucial aspect of statistical analysis.
These results indicate that BRS-DR is a robust and reliable method for estimating the ATE in various settings.
An essential practical contribution of this study is that BRS-DR could be considered the first choice for researchers seeking to estimate the ATE within a DR framework, especially when the underlying data generating process is uncertain.

\section*{Acknowledgement}
This work is supported by the Japan Society for the Promotion of Science (JSPS KAKENHI) grant numbers 21H00699 and 20H00080.

\vspace{1cm}
\bibliographystyle{chicago}
\bibliography{ref}

\newpage
\setcounter{equation}{0}
\setcounter{section}{0}
\setcounter{table}{0}
\setcounter{thm}{0}
\setcounter{page}{1}
\renewcommand{\thesection}{S\arabic{section}}
\renewcommand{\theequation}{S\arabic{equation}}
\renewcommand{\thetable}{S\arabic{table}}
\renewcommand{\thethm}{S\arabic{thm}}

\vspace{1cm}
\begin{center}
{\LARGE
{\bf Supplementary Material for ``Ensemble Doubly Robust Bayesian Inference via Regression Synthesis"}
}
\end{center}

This Supplementary Material provides the proofs of the theorems presented in the paper, the posterior sampling algorithm, and additional simulation results.

\section{Proof}

\subsection{Proof of Theorem~1 }
Our goal is to construct a predictive distribution, $\hat{\pi}_{n}\left(x\right)=p(Z=1|\mathcal{D}_{n},X=x)$,
that is consistent with the target, $p^{*}\left(Z=1\left|X=x\right.\right)$,
where a consistent predictive distribution is defined as 
\[
\lim_{n\rightarrow\infty}P\left(\Big|p^{*}\left(Z=1\left|X\right.\right)-p\left(Z=1\left|\mathcal{D}_{n},X\right.\right)\Big|<\delta\right)=1,\ \textrm{for any }\delta>0,
\]
where $P\left(\cdot\right)$ is a probability measure of the data
generating process.

Denote the realization of the random variable of the covariate, $X_{i}$,
as $y_{i}=y\left(X_{i}\right)$, $\delta_{0i}=\delta\left(X_{i}\right)$,
$z_{i}=Z\left(X_{i}\right)$, $\pi_{ji}=\pi_{j}\left(X_{i}\right)$,
$\delta_{ji}^{*}=\delta_{j}^{*}\left(X_{i}\right)$. Then, we have
\[
p\left(\left.Z_{i}\right|X_{i},\left\{ \pi_{j}\right\} _{j=1,\cdots,J}\right)=\textrm{Ber}\left(\frac{\exp\left(\delta_{0i}+\sum_{j=1}^{J}\delta_{ji}\pi_{ji}\right)}{1+\exp\left(\delta_{0i}+\sum_{j=1}^{J}\delta_{ji}\pi_{ji}\right)}\right).
\]
Denote the $n$ samples using the superscript: $\boldsymbol{z}^{n}=\left(z_{1},\cdots,z_{n}\right)^{\top}$,
$\boldsymbol{\delta}_{j}^{n}=\left(\delta_{j1},\cdots,\delta_{jn}\right)^{\top}$,
$\boldsymbol{\pi}_{j}^{n}=\left(\pi_{j1},\cdots,\pi_{jn}\right)^{\top}$.
$\delta^{*}\left(x\right)\in L^{2}\left(X\right),j=0,\cdots,J$ and
the parameter space, $L^{2}\left(X\right)$, is complete and separable;
i.e., a standard measure space. 

We now construct the predictive distribution, using a Gaussian process.
Let the prior distributions be $\delta_{0}\sim\textrm{GP}\left(\tau_{0},h_{0}\right)$
and $\boldsymbol{\delta}_{j}\sim\textrm{GP}\left(\tau_{j},h_{j}\right)$.
Here, we assume that $\boldsymbol{\delta}_{j}$ are independent for
$j$. The conditional likelihood function of $\boldsymbol{z}^{n}$,
given $\{\boldsymbol{\pi}_{j}^{n}\}_{j=1\cdots J}$ and $\mathcal{L}(\boldsymbol{z}^{n}|\boldsymbol{\delta}^{n},\{\boldsymbol{\pi}_{j}^{n}\}_{j=1\cdots J})$,
is 
\[
\mathcal{L}(\boldsymbol{z}^{n}|\boldsymbol{\delta}^{n},\{\boldsymbol{\pi}_{j}^{n}\}_{j=1\cdots J})=\prod_{i=1}^{n}\frac{\exp\left(\delta_{0}(\boldsymbol{x}_{i})+\sum_{j=1}^{J}\delta_{j}(\boldsymbol{x}_{i})\pi_{j}(\boldsymbol{x}_{i})\right)^{z_{i}}}{1+\exp\left(\delta_{0}(\boldsymbol{x}_{i})+\sum_{j=1}^{J}\delta_{j}(\boldsymbol{x}_{i})\pi_{j}(\boldsymbol{x}_{i})\right)},
\]
where the $\circ$ operator is the Hadamard product. The likelihood
function $\mathcal{L}\left(\boldsymbol{z}^{n}\left|\boldsymbol{\delta}^{n}\right.\right)$,
is convoluted by the agent density, $\left\{ h_{j}\right\} $, which
gives us 
\[
\mathcal{L}(\boldsymbol{z}^{n}|\boldsymbol{\delta}^{n})=\int_{\left[0,1\right]^{\otimes J}}\prod_{i=1}^{n}\frac{\exp\left(\delta_{0}(\boldsymbol{x}_{i})+\sum_{j=1}^{J}\delta_{j}(\boldsymbol{x}_{i})\pi_{j}(\boldsymbol{x}_{i})\right)^{z_{i}}}{1+\exp\left(\delta_{0}(\boldsymbol{x}_{i})+\sum_{j=1}^{J}\delta_{j}(\boldsymbol{x}_{i})\pi_{j}(\boldsymbol{x}_{i})\right)}\prod_{j=1}^{J}h_{j}\left(\left.\boldsymbol{\pi}_{j}^{n}\right|\boldsymbol{x}^{n}\right)d\pi_{j}.
\]
Then, the conditional joint probability density of $\boldsymbol{z}^{n+1}=((\boldsymbol{z}^{n})^{\top},z_{n+1})^{\top}$
given data set $\mathcal{D}^{n}=\left\{ \boldsymbol{z}^{n},\boldsymbol{x}^{n}\right\} $
is 
\begin{alignat}{1}
P\left(\left.\boldsymbol{z}^{n+1}\right|\mathcal{D}^{n}\right)= & \int_{\mathbb{R}^{J+1}}\mathcal{L}(\boldsymbol{z}^{n+1}|\boldsymbol{\delta}^{n+1})\varPi\left(\boldsymbol{\delta}^{n+1}\right)d\boldsymbol{\delta}^{n+1}\label{eq:Predictive}
\end{alignat}
and the predictive distribution of $z_{n+1}$ is 
\[
p\left(z_{n+1}\left|\mathcal{D}^{n},x_{n+1}\right.\right)=\frac{P\left(\left.\boldsymbol{z}^{n+1}\right|{x}^{n+1}\right)}{\sum_{z_{n+1}=1,2}P\left(\left.\boldsymbol{z}^{n+1}\right|{x}^{n+1}\right)}.
\]
In what follows, we show that the above predictive distribution is
consistent with the true distribution $p^{*}\left(z_{n+1}\left|\boldsymbol{x}_{n+1}\right.\right)$.

Assume, $0<p^{*}\left(Z=1\left|X=x\right.\right)$ for all $x$. We
show a stronger result, 
\begin{equation}
\left|p^{*}\left(Z=1\left|X\right.\right)-p\left(Z=1\left|\mathcal{D}^{n},X\right.\right)\right|\rightarrow0,\ \textrm{in probability }P.\label{eq:Consistency}
\end{equation}

Let the direct product measure (cylinder measure) of the data, $\left(z_{1},\cdots,z_{n},\cdots\right)$,
be $\mathbb{P}_{\infty}^{*}$, and the direct product measure (cylinder
measure) of the data of the synthesis model, $\left(z_{1},\cdots,z_{n},\cdots\right)$,
be $\mathbb{P}_{\infty}=\mathcal{L}\left(\boldsymbol{z}^{\infty}\left|\boldsymbol{\delta}^{\infty}\right.\right)$.
Denote the marginal under the Gaussian process prior as 
\[
\mathbb{P}_{\infty}\pi^{\otimes\infty}=\int_{\mathbb{R}^{J\otimes\infty}}\mathcal{L}\left(\boldsymbol{z}^{\infty}\left|\boldsymbol{\delta}^{\infty}\right.\right)\pi\left(\boldsymbol{\delta}^{\infty}\right)d\boldsymbol{\delta}^{\infty}.
\]
As with $\mathbb{P}_{n}^{*},\mathbb{P}_{n}\pi^{\otimes n}$, denote
the conditional distribution of the cylinder measure, given $\boldsymbol{z}^{n}$,
as $\left.\mathbb{P}_{\infty}^{*}\right|_{\boldsymbol{z}^{n}},\left.\mathbb{P}_{\infty}\pi^{\otimes\infty}\right|_{\boldsymbol{z}^{n}}$.
Note that 
\[
\left.\mathbb{P}_{\infty}^{*}\right|_{\boldsymbol{z}^{n}}=\frac{\mathcal{L}\left(\boldsymbol{z}^{\infty}\left|\boldsymbol{\delta}^{*\infty}\right.\right)}{\sum_{z_{n+1}}\cdots\sum_{z_{\infty}}\mathcal{L}\left(\boldsymbol{z}^{\infty}\left|\boldsymbol{\delta}^{*\infty}\right.\right)}
\]
and 
\[
\left.\mathbb{P}_{\infty}\pi^{\otimes\infty}\right|_{\boldsymbol{z}^{n}}=\frac{\int_{\mathbb{R}^{J\otimes\infty}}\mathcal{L}\left(\boldsymbol{z}^{\infty}\left|\boldsymbol{\delta}^{\infty}\right.\right)\pi\left(\boldsymbol{\delta}^{\infty}\right)d\boldsymbol{\delta}^{\infty}}{\sum_{z_{n+1}}\cdots\sum_{z_{\infty}}\int_{\mathbb{R}^{J\otimes\infty}}\mathcal{L}\left(\boldsymbol{z}^{\infty}\left|\boldsymbol{\delta}^{\infty}\right.\right)\pi\left(\boldsymbol{\delta}^{\infty}\right)d\boldsymbol{\delta}^{\infty}}.
\]
Given this notation, to prove (\eqref{eq:Consistency}), we need to
show, 
\begin{equation}
\left|\left.\mathbb{P}_{\infty}^{*}\right|_{\boldsymbol{z}^{n}}-\left.\mathbb{P}_{\infty}\pi^{\otimes\infty}\right|_{\boldsymbol{z}^{n}}\right|\rightarrow0,\ \textrm{in }\mathbb{P}_{\infty}^{*}.\label{eq:Consis-2}
\end{equation}

Now, since we assumed $0<\pi\left(\left\{ \boldsymbol{\delta}_{j}^{*}\right\} _{j=0,\cdots,J}\right)$,
$\mathbb{P}_{\infty}^{*}$ is absolute continuous with regard to the
marginal, $\mathbb{P}_{\infty}\pi^{\otimes\infty}$: $\mathbb{P}_{\infty}^{*}\ll\mathbb{P}_{\infty}\pi^{\otimes\infty}$.
Denote the likelihood ratio of $\mathbb{P}_{\infty}^{*}$ and $\mathbb{P}_{\infty}\pi^{\otimes\infty}$,
and its $\mathcal{D}_{n}$-conditional likelihood ratio as 
\[
\varDelta=\frac{d\mathbb{P}_{\infty}^{*}}{d\mathbb{P}_{\infty}\pi^{\otimes\infty}},\quad\varDelta_{n}=\mathbb{E}^{\mathbb{P}_{\infty}\pi^{\otimes\infty}}\left[\left.\frac{d\mathbb{P}_{\infty}^{*}}{d\mathbb{P}_{\infty}\pi^{\otimes\infty}}\right|\mathcal{D}_{n}\right].
\]
Here, $\mathbb{E}^{\mathbb{P}_{\infty}\pi^{\otimes\infty}}\left[\cdot\right]$
is an integral regarding the cylinder measure, $\mathbb{P}_{\infty}\pi^{\otimes\infty}$.
Further, since $\mathbb{E}^{\mathbb{P}_{\infty}\pi^{\otimes\infty}}\left[\varDelta_{n}\right]\leqq1$
for all $n\geqq0$, and from Doob's martingale convergence theorem,
we have 
\begin{equation}
\lim_{n\rightarrow\infty}\mathbb{E}^{\mathbb{P}_{\infty}\pi^{\otimes\infty}}\left[\left|\varDelta_{n}-\varDelta\right|\right]=0.\label{eq:DMCT}
\end{equation}
From this, we can prove (\eqref{eq:Consis-2}). For the function,
$h$, on $\mathbb{R}^{\infty}$, we have 
\begin{alignat*}{1}
 & \sup_{\left\Vert h\left(\cdot,z^{n}\right)\right\Vert \leqq1}\left|\int_{t^{n}}\left\{ \int h\left(\boldsymbol{z}^{n+1:\infty},\boldsymbol{z}^{n}\right)\left(\left.\mathbb{P}_{\infty}^{*}\right|_{\boldsymbol{z}^{n}}-\left.\mathbb{P}_{\infty}\pi^{\otimes\infty}\right|_{\boldsymbol{z}^{n}}\right)d\boldsymbol{z}^{n+1:\infty}\right\} p^{*}\left(\boldsymbol{z}^{n}\right)d\boldsymbol{z}^{n}\right|\\
= & \sup_{\left\Vert h\left(\cdot,z^{n}\right)\right\Vert \leqq1}\left|\int_{t^{n}}\int h\left(\boldsymbol{z}^{n+1:\infty},\boldsymbol{z}^{n}\right)\left\{ Z-Z_{n}\right\} \left.\mathbb{P}_{\infty}\pi^{\otimes\infty}\right|_{\boldsymbol{z}^{n}}d\boldsymbol{z}^{n+1:\infty}\mathbb{P}_{n}\pi^{\otimes n}d\boldsymbol{z}^{n}\right|\\
\leqq & \mathbb{E}^{\mathbb{P}_{\infty}\pi^{\otimes\infty}}\left[\left|Z_{n}-Z\right|\right]\rightarrow0.
\end{alignat*}
If we let $h\left(\cdot,z^{n}\right)=1_{A}\left(\cdot\right)$ where
$A\in\mathcal{B}\left(\left\{ 0,1\right\} ^{\infty}\right)$, we obtain
the result.

\section{Detailed Sampling Steps of the MCMC Algorithm}
\label{sec:org447e88c}

\subsection{BRS for continuous response}

We introduce the notational simplifications \(\beta_{ji} = \beta_j(X_i)\), \(f_{ji} = f_j(X_i)\),
\(\boldsymbol{y} = (y_1, \dots, y_n)^{\top}\) and \(\mathbf{X} = (X_1, \dots, X_n)^{\top}\).
We also define \(\Phi\) as a set of varying coefficients and latent variables, encompassing
\(\Phi = \{\beta_{0i}, \dots, \beta_{Ji}, f_{1i}, \dots, f_{Ji}\}_{i=1, \dots, n}\).
Note that the prior distributions for the unknown parameters as $\tau_j\sim {\rm IG}(\delta_j/2, \xi_j/2)$, $\psi_j\sim U(\underline{c}_\psi, \overline{c}_\psi)$ and $\sigma^2\sim {\rm IG}(\delta_\sigma/2, \xi_\sigma/2)$.
In what follows, we assume that
\(f_{ji} \sim N (a_{ji}, b_{ji})\) independently for \(j=1, \dots, J\) and \(i=1, \dots, n\), where \(a_{ji}\) and \(b_{ji}\) are provided by the \(j\)-th model (agent) and are fixed values.
The joint posterior distribution of \(\theta\) and \(\Phi\) is given by
\begin{equation}
\label{eq:jointPosteriorGauss}
\begin{aligned}
p( \theta, \Phi | \boldsymbol{y}, \mathbf{X})
& \propto
\pi(\theta)
\prod_{i=1}^n \phi \Big(Y_i; \beta_{0i} + \sum_{j=1}^J \beta_{ji} f_{ji}, \sigma^2 \Big)
\prod_{j=1}^J h_j(f_{ji})\prod_{j=0}^J \phi_n(\beta_j; \bar{\beta}_j, \tau_j^2 H_m(\psi_j ; \mathcal{X})) \\
\end{aligned}
\end{equation}
where
\(\phi\) is a density function of a normal distribution,
\(\phi_n\) is a density function of a multivariate normal distribution,
\(\beta_j = (\beta_{j1}, \dots, \beta_{jn})\) and
\(\pi(\theta)\) is a joint prior distribution of \(\theta\).
Note that the multivariate normal distribution $\phi_n(\beta_j; \bar{\beta}_j, \tau_j^2 H_m(\psi_j ; \mathcal{X}))$ induced by \(m\)-nearest neighbor Gaussian process can be expressed as
\begin{equation}
\label{eq:betaPrior}
\phi_n(\beta_j; \bar{\beta}_j, \tau_j^2 H_m(\psi_j ; \mathcal{X}))= \prod_{i=1}^n \phi(\beta_j^{\ast}(X_i); B_j(X_i) \beta_j^{\ast}(N(X_i)), \tau_j^2 F_j(X_i)),
\end{equation}
with
\(\beta_j^{\ast}(X_i) = \beta_j(X_i) - \bar{\beta}_j\), where
\begin{align*}
B_j(X_i) &= C_j(X_i, N(X_i); \psi_j) C_j(N(X_i), N(X_i); \psi_j)^{-1}, \\
F_j(X_i) &= 1 - C_j(X_i, N(X_i); \psi_j) C_j(N(X_i), N(X_i); \psi_j)^{-1} C_j(N(X_i), X_i; \psi_j),
\end{align*}
and \(N(X_i)\) represents an index set of \(m\)-nearest neighbors of \(X_i\).
Here, \(C_j(t,v;\psi_j)\) for \(t=(t_1, \dots, t_{d_t}) \in \mathbb{R}^{d_t}\) and
\(v=(v_1, \dots, v_{d_v}) \in \mathbb{R}^{d_v}\) denotes a \(d_t \times d_v\) correlation matrix whose \((i_t, i_v)\)-element is \(C(||t_{it}-v_{iv}||;\psi_j)\).
The Gibbs sampling algorithm to sample from (\ref{eq:jointPosteriorGauss}) is described as follows:

\begin{itemize}
\item[-]
{\bf (Sampling of $\beta_{ji}$)} \ \
The full conditional distribution of $\boldsymbol{\beta}_i = (\beta_{0i}, \dots, \beta_{Ji})$ is $N(\bar{\beta}_j + A_i^{-1} B_i, A_i^{-1})$, where
\begin{align*}
A_i &= \frac{f_i f_i^{\top}}{\sigma^2} + \text{diag} \left( \gamma_{0i}, \dots, \gamma_{Ji} \right), \ \ \ \ \
\gamma_{ji} = \frac{1}{\tau_j^2 F_j(X_i)}
+\sum_{t;X_i \in N(t)} \frac{ B_j(t;X_i)^2 }{ \tau_j^2 F_j(t) }\\
B_i &= \frac{f_i Y_i}{\sigma^2} + (m_{0i}, \dots, m_{Ji})^{\top},
\end{align*}
and
\begin{align*}
m_{ji} &= \frac{ B_j(X_i)^{\top} \beta_j^{\ast}(N(X_i)) }{ \tau_j^2 F_j(X_i) }
+\sum_{t;X_i \in N(t)} \frac{ B_j(t;X_i) }{ \tau_j^2 F_j(t) }
\left\{ \beta_j^{\ast}(t) - \sum_{s \in N(t), s \neq X_i} B_j(t;s) \beta_j^{\ast}(s) \right\},
\end{align*}
and \(f_i = (1, f_{1i}, \dots, f_{Ji})^{\top}\).
Here, \(B_j(t;s)\) denotes the scalar coefficient for \(\beta_j^{\ast}(X_i)\) among the elements of the coefficient vector \(B_j(t)\).

\item[-]
{\bf (Sampling of $f_{ji}$)} \ \
The full conditional distribution of \(f_{ji}\) is given by $N((A_{ji}^{(f)})^{-1} B_{ji}^{(f)}, (A_{ji}^{(f)})^{-1})$, where
\begin{align*}
A_{ji}^{(f)} = \frac{ \beta_{ji}^2 }{\sigma^2} + \frac{1}{b_{ji}},  \ \ \ \ \
B_{ji}^{(f)} = \frac{ \beta_{ji} }{\sigma^2}
\bigg\{ (Y_i - \beta_{0i}) + \sum_{j \neq k} \beta_{ki} f_{ki} \bigg\} + \frac{a_{ji}}{b_{ji}}.
\end{align*}

\item[-]
{\bf (Sampling of $\tau_j^2$)} \ \ The full conditional distribution of \(\tau_j^2\) is given by
$$
{\rm IG}\left( \frac{\delta_j + n}{2},
\frac{ \xi_j }{2} + \frac{1}{2} \sum_{i=1}^n
\frac{ \big\{ \beta_j^{\ast}(X_i) - B_j(X_i) \beta_j^{\ast}(N(X_i)) \big\}^2 }{F_j(X_i)} \right).
$$

\item[-]
{\bf (Sampling of $\psi_j$)} \ \
The full conditional distribution of \(\psi_j\) is proportional to
\begin{align*}
I(\underline{c}_\psi\leq \psi_j\leq \overline{c}_\psi)\prod_{i=1}^n
\phi(\beta_j^{\ast}(X_i); B_j(X_i) \beta_j^{\ast}(N(X_i)),
\tau_j^2 F_j(X_i)),
\end{align*}
from which we use the random-walk Metropolis-Hastings algorithm to generate a random sample.

\item[-]
{\bf (Sampling of $\sigma^2$)} \ \
The full conditional distribution of \(\sigma^2\) is given by
$$
{\rm IG} \left( \frac{\delta_{\sigma} + n}{2},
\frac{ \eta_{\sigma} }{2} + \frac{1}{2} \sum_{i=1}^n
\Big( Y_i - \beta_0 - \sum_{j=1}^J \beta_{ji} f_{ji} \Big)^2 \right).
$$
\end{itemize}

\subsection{BRS for binary response}
Under the binary model $P(Y_i = 1 | \eta_i) =  \exp(\eta_i)/\{1 + \exp(\eta_i)\}$, the joint posterior distribution of $\theta$ and \(\Phi\) is given by
\begin{equation}
\label{eq:jointPosteriorBin}
\begin{aligned}
\pi(\theta, \Phi | \boldsymbol{y}, \mathbf{X})
&\propto \pi(\theta)\prod_{i=1}^n \frac{ \exp(Y_i\eta_i) }{ 1 + \exp(\eta_i) }
\prod_{j=1}^J h_j(f_{ji})
\prod_{j=0}^J \phi_n(\beta_j^{(s)}; \bar{\beta}_j, \tau_j^2 H_m(\psi_j ; \mathcal{X})),
\end{aligned}
\end{equation}
where $\eta_i = \beta_0(X_i) + \sum_{j=1}^J \beta_j(X_i) f_j(X_i)$.
Again, we assume that
\(f_{ji} \sim N (a_{ji}, b_{ji})\) is received independently for \(j=1, \dots, J\) and \(i=1, \dots, n\), where \(a_{ji}\) and \(b_{ji}\) are provided by the \(j\)th model (agent).
To derive a Gibbs sampler, we employ the Polya-gamma data augmentation technique \citep{Polson_etal2013PolyaGamma_JASA} for the binomial likelihood as
$$
 \frac{ \exp(Y_i\eta_i) }{ 1 + \exp(\eta_i) }
= \frac{1}{2} \exp(\kappa_i \eta_i)
\int_0^{\infty} \exp \left( -\frac{ \omega_i \eta_i^2 }{2} \right) p(\omega_i; 1,0) d\omega_i,
$$
where $\kappa_i=Y_i-1/2$ and $p(\omega_i; b,c)$ is the density of the Polya-gamma distribution with parameters $(b,c)$.
Using the expression (\ref{eq:betaPrior}), the Gibbs sampling algorithm is obtained as follows:

\begin{itemize}
\item[-]
{\bf (Sampling of $\beta_{ji}$)} \ \
The full conditional distribution of $\boldsymbol{\beta}_j$ is $N(\boldsymbol{A}_j \boldsymbol{D}_j, \boldsymbol{A}_j)$,
where
\begin{align*}
\boldsymbol{A}_j &= \left\{
\frac{\boldsymbol{F}_j}{\tau_j^2}
+\mathrm{diag}(\boldsymbol{\omega} \circ \boldsymbol{f}_j^2)
\right\}^{-1} \\
\boldsymbol{D}_j &=
\boldsymbol{f}_j \circ \left[
\boldsymbol{\kappa}
-\boldsymbol{\omega} \circ \left( \sum_{k \neq j} \boldsymbol{\beta}_k \circ \boldsymbol{f}_k \right) \right]
+\left( \tau_j^2 \right)^{-1} \boldsymbol{F}_j \left( \bar{\beta}_j + \boldsymbol{B}_j \boldsymbol{\beta}_j^{\ast} \right).
\end{align*}
Here \(\boldsymbol{\kappa} = (\kappa_1, \dots, \kappa_n)^{\top}\), $\boldsymbol{F}_j={\rm diag}(F_j(X_1),\ldots,F_j(X_n))^\top $
\(\boldsymbol{f}_j^2 = (f_{j1}^2, \dots, f_{jn}^2)^{\top}\) and \(\circ\) represents the Hadamard product.

\item[-]
{\bf (Sampling of $\omega_i$)} \ \
The full conditional distribution of $\omega_i$ is ${\rm PG}(1,  \beta_{0i} + \sum_{j=1}^J \beta_{ji} f_{ji})$.

\item[-]
{\bf (Sampling of $f_{ji}$)} \ \
The full conditional distribution of \(f_{ji}\) is given by $N((A_{ji}^{(f)})^{-1}B_{ji}^{(f)}, (A_{ji}^{(f)})^{-1})$, where
\begin{align*}
A_{ji}^{(f)}= \omega_i \beta_{ji}^2 + \frac{1}{b_{ji}}, \ \ \ \
B_{ji}^{(f)}= \kappa_i \beta_{ji} - \frac{\omega_i \beta_{ji}}{2} \left( \beta_{0i} + \sum_{j \neq k} \beta_{ki} f_{ki} \right) - \frac{a_{ji}}{b_{ji}}.
\end{align*}

\item[-]
{\bf (Sampling of $\tau_j^2$ and $\psi_j$)} \ \ The full conditional distributions of these parameters are the same as those in the continuous response.
\end{itemize}

\section{Additional simulation results}
\label{sec:org64fd386}
\renewcommand{\arraystretch}{0.8}
\renewcommand{\doublerulesep}{1pt}

\subsection{Simulation 3: Correct and incorrect specifications}
\label{sec:sim3}
We here evaluate the effect of model specification/misspecification on the doubly robust estimators and the improvement by combining the multiple estimators.
The data generating process, drawing inspiration from \citet{Saarela2016BayesianDR_Biometrika} and extensively employed in various studies, is structured as follows:
\begin{align*}
&X_1, X_2, X_3, X_4 \sim N (0, 1), \ \ \ \
U_1= \frac{|X_1|}{ \sqrt{1 - 2/\pi} } \\
Z &| U_1, X_2, X_3 \sim \text{Ber} \{ \text{logistic}(0.4 U_1 + 0.4 X_2 + 0.8 X_3)\}, \\
Y &| Z, U_1, X_2, X_4 \sim N ( Z - U_1 - X_2 - X_4, 1 ).
\end{align*}
This data generating process shows that the true ATE equals the individual treatment effect for each unit, which is \(1\).
We considered three cases of sample sizes, $n = 200, 1000$ and $2000$.

For the simulated dataset, we applied the following three methods:
\begin{itemize}
\item M1: DR estimator using misspecified logistic/linear regression with covariates \(\exp(X_1), x_2, x_3, x_4\).
\item M2: DR estimator using misspecified logistic/linear regression with covariates \(X_1^3, x_2, x_3, x_4\).
\item M3: DR estimator using four types of covariates, given as follows:
\begin{itemize}
\item Scenario 1: correctly specified logistic/linear regression with covariates
\(u_1, x_2, x_3, x_4\).
\item Scenario 2: correctly specified logistic regression with covariates
\(u_1, x_2, x_3\) and misspecified linear regression with covariates
\(X_1, x_2, x_4\).
\item Scenario 3: misspecified logistic regression with covariates
\(X_1, x_2, x_3\) and correctly specified linear regression with covariates
\(u_1, x_2, x_4\).
\item Scenario 4: misspecified logistic/linear regression with covariates
\(X_1, x_2, x_3, x_4\).
\end{itemize}
\end{itemize}
Furthermore, we combine M1, M2 and M3 through SA, SIC, BMA and BRS as used in Section~\ref{sec:sim1}.
We repeated 500 times to compute the estimator of ATE to quantify the bias, defined as the average discrepancy between the true ATE and the estimates,  and the mean squared error (MSE), reflecting the average of the squared differences.
Furthermore, the coverage probabilities (CP) and the average lengths (AL) of the 95\% confidence/credible intervals for the ATE are documented in Table \ref{tab:sim3:CpAl}.

\begin{figure}[htbp]
\includegraphics[width=0.95\textwidth]{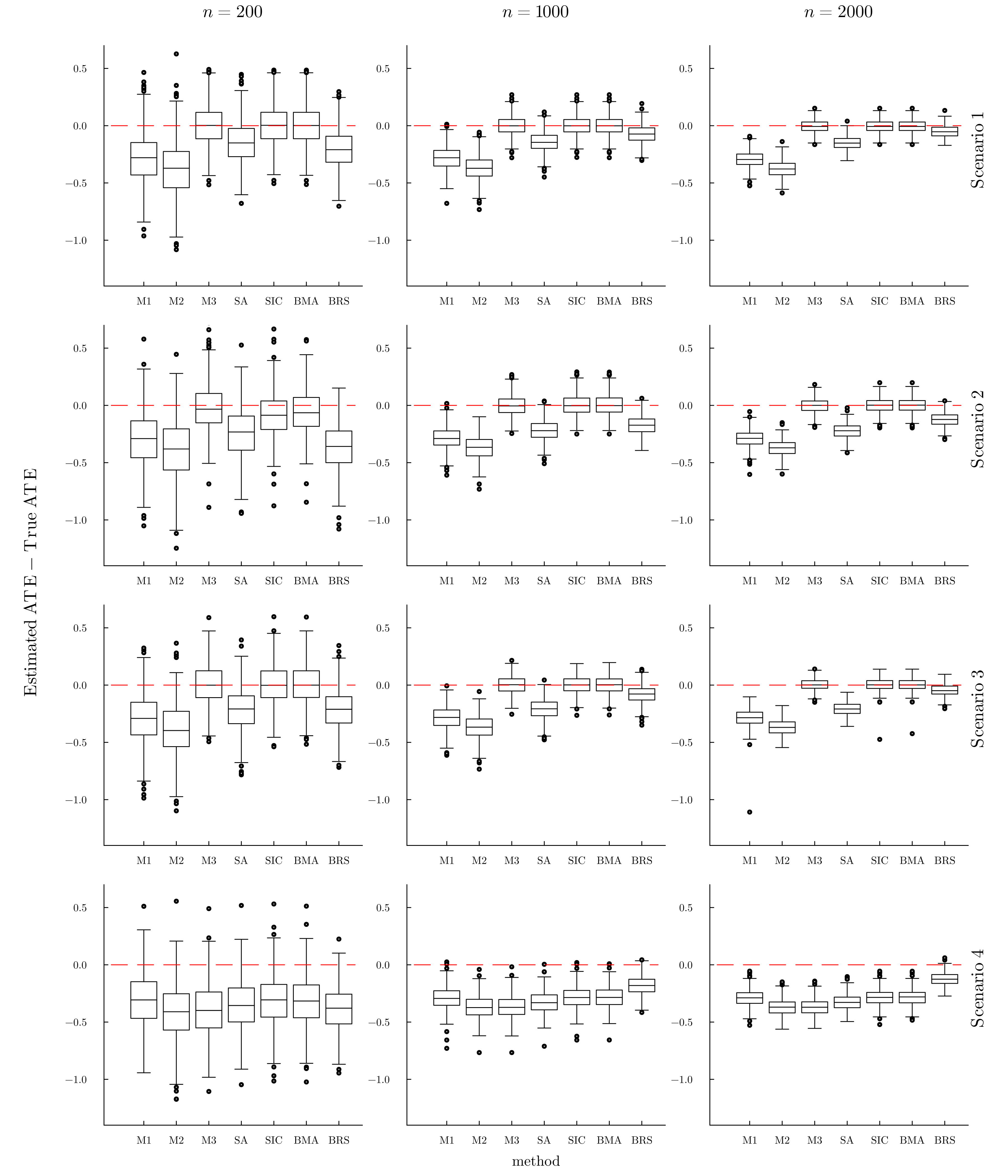}
\caption{\label{fig:sim3:errorbox}
Boxplots of point estimates of the ATE based on 500 replications under Simulation 3.}
\end{figure}

From Figure~\ref{fig:sim3:errorbox}, we can observe that when the correct model was included in either propensity score or outcome estimation (Scenarios 1, 2, and 3),
the third model (model 3) exhibited notable success in achieving unbiased estimation, aligning with the theoretical expectations of DR estimators.
SIC and BMA also tended to successfully converge towards the third model, apparently achieving unbiased estimation.
In contrast, the performance of SA was not as effective.
Although not the most effective with smaller sample sizes, BRS displayed considerable improvement as the sample size grew, closing the gap with the best-performing method.
Such observations indicate that when prior weights are equal for all models, BRS-DR may not converge to the correct model as quickly as SIC and BMA but does so eventually with an increase in sample size.
In Scenario 4, where all propensity score and outcome models were misspecified, BRS notably outperformed other methods, especially with larger sample sizes.
Regarding CP, other methods showed significantly lower values, between 0.2\% and 16.0\%, while BRS maintained a relatively high coverage probability, around 50\%.
These findings imply that BRS's ability to measure uncertainty remains more robust against model misspecifications than other ensemble techniques.

\begin{table}
\caption{\label{tab:sim3:CpAl}
Coverage probability (CP) and average length (AL) of 95\% confidence/credible intervals of estimated ATE for different scenarios and sample sizes, averaged over 500 replications, under Simulation 3.}
    \centering
    \scalebox{0.95}{
    \begin{tabular}{ccrrrrrrr}
    \toprule
    \multicolumn{9}{c}{CP (\%)} \\
    \textit{n} & Scenario & \multicolumn{1}{c}{M1} & \multicolumn{1}{c}{M2} & \multicolumn{1}{c}{M3} & \multicolumn{1}{c}{SA} & \multicolumn{1}{c}{SIC} & \multicolumn{1}{c}{BMA} & \multicolumn{1}{c}{BRS} \\ \midrule
        200 & 1 & 70.8 & 61.0 & 93.8 & 87.2 & 93.6 & 93.6 & 94.6 \\
            & 2 & 69.6 & 57.0 & 98.0 & 76.8 & 93.8 & 95.4 & 78.0 \\
            & 3 & 71.8 & 57.8 & 91.0 & 75.4 & 90.4 & 90.4 & 93.2 \\
            & 4 & 68.4 & 54.2 & 56.0 & 62.2 & 68.8 & 66.6 & 77.2 \\ \midrule
        1000 & 1 & 18.8 & 7.6 & 93.2 & 57.6 & 92.6 & 93.0 & 90.2 \\
             & 2 & 17.0 & 4.8 & 99.0 & 39.4 & 97.4 & 97.2 & 54.8 \\
             & 3 & 18.2 & 4.8 & 94.0 & 29.8 & 94.0 & 94.2 & 88.4 \\
             & 4 & 14.4 & 4.4 & 4.8 & 7.2 & 16.0 & 15.4 & 50.2 \\ \midrule
        2000 & 1 & 1.2 & 0.0 & 94.4 & 29.4 & 94.6 & 94.8 & 90.2 \\
             & 2 & 0.8 & 0.0 & 98.6 & 8.8 & 97.4 & 97.4 & 50.4 \\
             & 3 & 1.0 & 0.0 & 96.6 & 5.6 & 96.0 & 95.8 & 95.0 \\
             & 4 & 1.4 & 0.2 & 0.4 & 0.8 & 1.4 & 1.6 & 48.2 \\ \bottomrule
    \multicolumn{9}{c}{AL} \\
    \textit{n} & Scenario & \multicolumn{1}{c}{M1} & \multicolumn{1}{c}{M2} & \multicolumn{1}{c}{M3} & \multicolumn{1}{c}{SA} & \multicolumn{1}{c}{SIC} & \multicolumn{1}{c}{BMA} & \multicolumn{1}{c}{BRS} \\ \midrule
        200 & 1 & 0.85 & 0.91 & 0.66 & 0.73 & 0.65 & 0.66 & 0.98 \\
            & 2 & 0.85 & 0.90 & 1.00 & 0.84 & 0.88 & 0.89 & 1.03 \\
            & 3 & 0.85 & 0.90 & 0.65 & 0.74 & 0.65 & 0.65 & 0.98 \\
            & 4 & 0.85 & 0.92 & 0.90 & 0.86 & 0.85 & 0.84 & 1.03 \\ \midrule
        1000 & 1 & 0.38 & 0.39 & 0.30 & 0.33 & 0.30 & 0.30 & 0.35 \\
             & 2 & 0.38 & 0.39 & 0.46 & 0.38 & 0.43 & 0.43 & 0.36 \\
             & 3 & 0.38 & 0.39 & 0.29 & 0.33 & 0.29 & 0.29 & 0.35 \\
             & 4 & 0.37 & 0.39 & 0.39 & 0.38 & 0.37 & 0.37 & 0.36 \\ \midrule
        2000 & 1 & 0.27 & 0.28 & 0.21 & 0.23 & 0.21 & 0.21 & 0.23 \\
             & 2 & 0.27 & 0.28 & 0.33 & 0.27 & 0.30 & 0.30 & 0.24 \\
             & 3 & 0.27 & 0.28 & 0.20 & 0.23 & 0.21 & 0.21 & 0.23 \\
             & 4 & 0.29 & 0.28 & 0.28 & 0.27 & 0.29 & 0.28 & 0.24 \\ \bottomrule
    \end{tabular}
    }
\end{table}

\subsection{Simulation 4: Mixture of different types of functional forms}
\label{sec:sim4}
Finally, we explore a scenario where the actual structure of the propensity score and outcome models diverges based on a particular covariate.
This approach mirrors the complexities inherent in empirical datasets, wherein a single model may demonstrate optimal fit for a distinct subgroup while other models provide superior alignment for different subgroups.
Through this simulation, we endeavor to evaluate the performance of BRS-DR within a multifaceted context that more accurately represents the heterogeneity of real-world data.

We first generate the covariate matrix as $(X_1, X_2, X_3, X_6,\dots, X_q)^{\top} \sim N_{q-2}(\boldsymbol{0}, \boldsymbol{I}_{q-2})$, $X_4 \sim \text{Ber}(p_4)$ with $p_4 \sim U(0, 1)$, and
$$
X_5 \sim {\rm MN}\left(\frac{p_{5,1}}{p_{5,1}+p_{5,2}+p_{5,3}}, \frac{p_{5,2}}{p_{5,1}+p_{5,2}+p_{5,3}}, \frac{p_{5,3}}{p_{5,1}+p_{5,2}+p_{5,3}}\right),
$$
with $p_{5,j} \sim U(0, 1)$ for $j = 1, 2, 3$.
The data-generating process for the treatment assignment is described as
\begin{align*}
&P(T=1 | X) = \begin{cases}
\text{logistic}(-0.5 + 0.3 X_1 + 0.5 X_2 X_4 + 0.6 X_3) \text{ if } X_5 = 1, \\
\text{logistic}(-0.5 + 0.3 X_1^2 + 0.5 X_2 X_4 + 0.6 X_3^2) \text{ if } X_5 = 2, \\
\text{logistic}(-0.5 + 0.3 \exp(X_1) + 0.5 X_2 X_4 + 0.6 |X_3|) \text{ if } X_5 = 3.
\end{cases}
\end{align*}
We then generate the outcome as $Y = \mu_0(X) + Z\tau(X)+ \epsilon$ with $\epsilon\sim N (0, 1)$, where
\begin{align*}
&\mu_0(X) = \begin{cases}
-7 + 6 X_3 \text{ if } X_5 = 1, \\
2 + 2 X_3^2 \text{ if } X_5 = 2, \\
2 + 2 \sin(3X_3) \text{ if } X_5 = 3, \\
\end{cases} \\
&\tau(X) = 1 + 2 X_2 X_5 + \frac{X_3^2}{2}.
\end{align*}
The complexity of this data-generating process is evident, as the structure for both the propensity score model and the outcome model is contingent on the ternary covariate \(X_5\).
For individuals with \(X_5 = 1\), a linear model will be most suitable for the propensity score and outcome modeling.
Conversely, a quadratic model will become more appropriate for individuals with \(X_5 = 2\).
Consistent with previous simulations, we investigate six combinations of sample sizes and numbers of covariates (\(n = 200, 1000, 2000\) and \(q = 5, 100\)).
We adopted the same method for ATE estimation as in Section~\ref{sec:sim1}.
Variable selection is performed similarly to the prior simulation for the high-dimensional case where \(q=100\).

\begin{figure}[htbp!]
\includegraphics[width=1.0\textwidth]{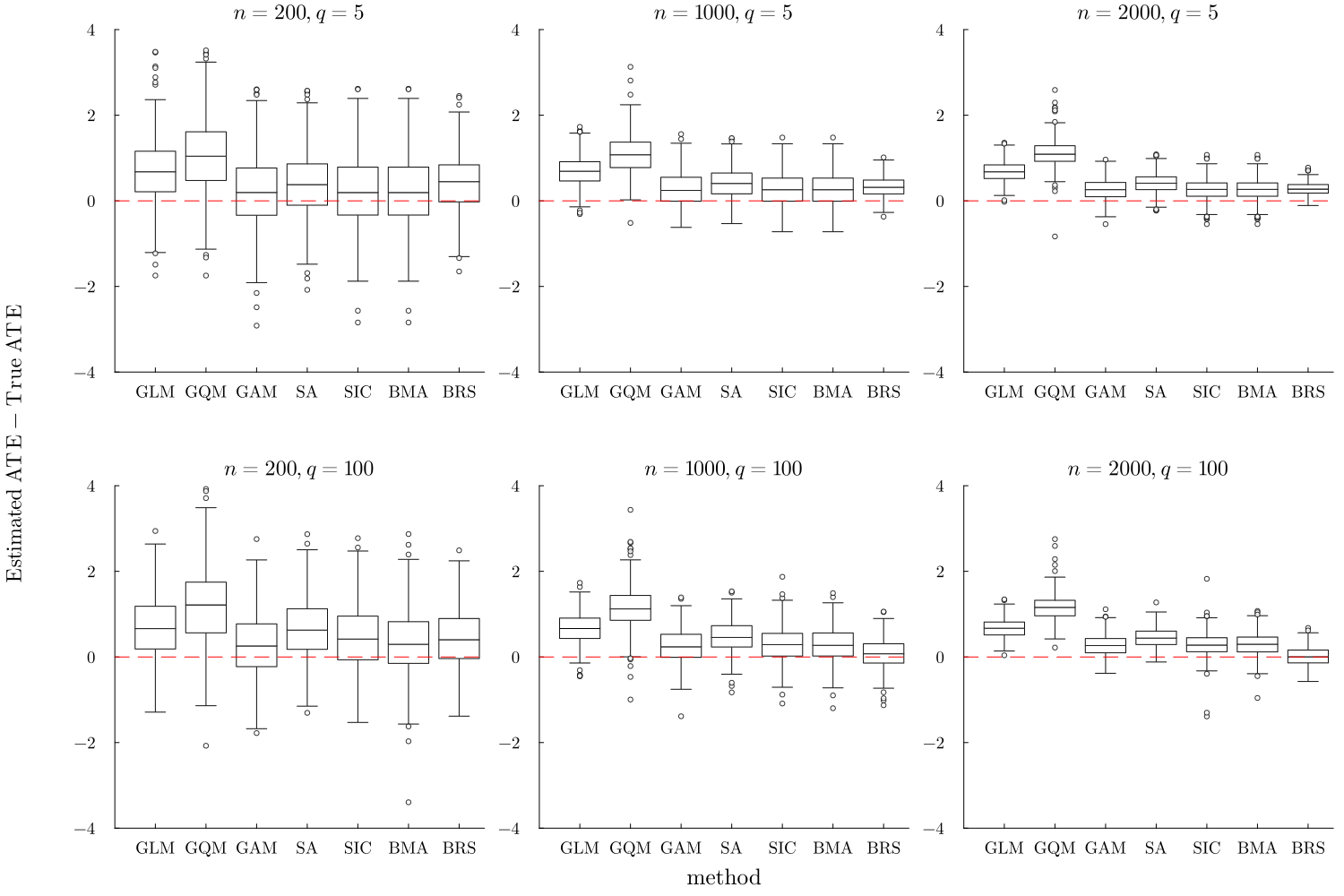}
\caption{\label{fig:sim4:errorbox}
Boxplots of difference of point estimates of the ATE based on 500 replications under Simulation~4 with $X_3$ observed (upper) and omitted (lower).}
\end{figure}

In Figure \ref{fig:sim4:errorbox}, we present the boxplots of the estimation error of comparative methods under 500 Monte Carlo replications.
We can see that the biases and variances are generally elevated compared to those in the previous simulation.
Such increases seem to result from the more intricate data-generating process within this simulation, presenting a more significant challenge for the models to discern and accurately represent the genuine underlying relationships.
BRS exhibits low variances, paralleling its performance in the prior simulation.
However, the estimates of BRS exhibit higher bias at smaller sample sizes (\(n=200\)), with the extent of bias diminishing as the sample size expands.
This trend mirrors observations from the previous simulation.
Notably, in sparser settings (\(q=100\)) with sufficient sample sizes (\(n=1000, 2000\)), the performance of BRS is particularly impressive.

Furthermore, the performance of 95\% credible/confidence intervals of ATE are shown in Table~\ref{tab:sim4:CpAl}.
The coverage probabilities for BRS approximate the ideal 95\% mark across most scenarios, reinforcing its reliability in diverse analytical contexts.
The simulation results demonstrate BRS-DR's proficiency in effectively handling complex settings.
When single models and other ensemble methods struggle to accurately estimate the ATE and its associated uncertainty, BRS-DR maintains its capability to deliver reliable results.

\begin{table}
\caption{\label{tab:sim4:CpAl}
Coverage probability (CP) and average length (AL) of 95\% confidence/credible intervals of estimated ATE for different scenarios and sample sizes, averaged over 500 replications under Simulation 4. }
    \centering
    \scalebox{0.95}{
    \begin{tabular}{cccrrrrrrr}
    \toprule
    \multicolumn{10}{c}{CP(\%)} \\
        Omission of $X_3$ & $n$ & $q$ & \multicolumn{1}{c}{GLM} & \multicolumn{1}{c}{GQM} & \multicolumn{1}{c}{GAM} & \multicolumn{1}{c}{SA} & \multicolumn{1}{c}{SIC} & \multicolumn{1}{c}{BMA} & \multicolumn{1}{c}{BRS} \\ \midrule
        No  & 200 & 4 & 89.4 & 82.2 & 75.8 & 83.8 & 80.2 & 80.6 & 95.8 \\
            &     & 100 & 80.0 & 54.8 & 62.0 & 60.4 & 71.6 & 72.6 & 88.0 \\
            & 1000 & 4 & 87.8 & 61.8 & 83.4 & 84.6 & 84.8 & 84.8 & 84.8 \\
            &      & 100 & 91.4 & 62.0 & 85.6 & 79.6 & 87.6 & 89.0 & 90.0 \\
            & 2000 & 4 & 86.6 & 50.4 & 86.8 & 86.4 & 86.8 & 87.0 & 85.2 \\
            &      & 100 & 90.4 & 48.4 & 87.4 & 83.8 & 89.2 & 89.0 & 93.8 \\ \midrule
        Yes & 200 & 4 & 88.2 & 79.6 & 81.2 & 87.2 & 83.4 & 83.0 & 98.2 \\
            &     & 100 & 82.8 & 62.2 & 73.0 & 76.0 & 77.6 & 77.0 & 93.8 \\
            & 1000 & 4 & 79.2 & 58.4 & 76.4 & 89.4 & 76.4 & 76.8 & 94.0 \\
            &      & 100 & 76.6 & 53.8 & 74.8 & 89.0 & 74.6 & 73.2 & 94.2 \\
            & 2000 & 4 & 62.4 & 36.0 & 58.2 & 85.4 & 60.8 & 60.4 & 86.0 \\
            &      & 100 & 65.8 & 36.8 & 66.4 & 90.0 & 64.6 & 64.2 & 91.0 \\
    \bottomrule
    \multicolumn{10}{c}{AL} \\
        Omission of $X_3$ & $n$ & $q$ & \multicolumn{1}{c}{GLM} & \multicolumn{1}{c}{GQM} & \multicolumn{1}{c}{GAM} & \multicolumn{1}{c}{SA} & \multicolumn{1}{c}{SIC} & \multicolumn{1}{c}{BMA} & \multicolumn{1}{c}{BRS} \\ \midrule
        No  & 200 & 4 & 2.14 & 3.05 & 1.71 & 1.32 & 1.8 & 1.8 & 1.47 \\
            &     & 100 & 1.31 & 1.7 & 0.92 & 0.83 & 1.2 & 1.23 & 1.3 \\
            & 1000 & 4 & 1.1 & 2.73 & 1.06 & 0.77 & 1.07 & 1.07 & 0.62 \\
            &      & 100 & 1.36 & 2.23 & 1.04 & 0.69 & 1.27 & 1.29 & 0.59 \\
            & 2000 & 4 & 0.84 & 3.73 & 0.86 & 0.6 & 0.86 & 0.86 & 0.46 \\
            &      & 100 & 0.95 & 1.8 & 0.91 & 0.56 & 0.95 & 0.96 & 1.6 \\ \midrule
        Yes & 200 & 4 & 1.86 & 2.63 & 1.37 & 1.35 & 1.45 & 1.45 & 1.42 \\
            &     & 100 & 1.38 & 1.34 & 1.02 & 0.9 & 1.16 & 1.15 & 1.31 \\
            & 1000 & 4 & 0.9 & 1.59 & 0.85 & 0.72 & 0.83 & 0.83 & 0.58 \\
            &      & 100 & 0.94 & 1.42 & 0.81 & 0.63 & 0.8 & 0.81 & 0.54 \\
            & 2000 & 4 & 0.63 & 1.25 & 0.62 & 0.5 & 0.6 & 0.6 & 0.42 \\
            &      & 100 & 0.65 & 1.17 & 0.61 & 0.47 & 0.59 & 0.6 & 0.38 \\
    \bottomrule
    \end{tabular}
    }
\end{table}

\end{document}